# A molecular dynamics simulation study on the frustrated Lewis pairs in ionic liquids

Xiaoqing Liu,[a†] Xiaojing Wang,[a†] Tianhao Yu,[b] Weizhen Zhao,[b] Lei Liu,[b*]

[a] School of Chemical and Environmental Engineering, North University of China, Taiyuan, 030051, PR China

[b] Institute of Process Engineering, Chinese Academy of Sciences, Beijing 100049, China

Corresponding author:

Lei Liu, liulei@ipe.ac.cn; liulei3039@gmail.com;

† These authors contributed equally to the work

**Abstract**: Steric hindered frustrated Lewis pairs (FLPs) has been shown to activate hydrogen molecules, and their reactivity is strongly determined by the geometric parameters of the Lewis acids and bases. A recent experimental study showed that ionic liquids (ILs) could largely improve the effective configuration of FLPs. However, the detailed mechanistic profile is still unclear. Herein, we performed a molecular dynamics (MD) simulations, aiming to reveal the effects of ILs on the structures of FLPs, and to present a rule for selecting more efficient reaction media. For this purpose, mixture systems were adopt consisting of the ILs [$C_n$mim][$NTf_2$] ($n$ = 6, 10, 14), and the typical FLP ($t$Bu)$_3$P/B($C_6F_5$)$_3$. Radial distribution function (RDF) results show that toluene competes with ($t$Bu)$_3$P to interact with B($C_6F_5$)$_3$, resulting in a relatively low effective ($t$Bu)$_3$P/B($C_6F_5$)$_3$ complex. [$C_n$mim][$NTf_2$] is more intended to form a solvated shell surrounding the ($t$Bu)$_3$P/B($C_6F_5$)$_3$, which increases the amount of effective FLPs. Spatial distribution function (SDF) results show that toluene formed a continuum solvation shell, which hinders the interactions of ($t$Bu)$_3$P and B($C_6F_5$)$_3$, while [$C_n$mim][$NTf_2$] leave a relatively large empty space, which is accessible by ($t$Bu$_3$)P or ($t$Bu$_3$)P molecules, resulting in a higher probability of Lewis acids and bases interactions. Lastly, we find that the longer alkyl chain length of [$C_n$mim] cations, the higher probability of effective FLPs.

**Introduction**

On the one hand, Lewis acids (LA) and Lewis bases (LB) with large steric hindrance cannot react with each other to form the thermodynamically stable classic Lewis adducts[1]. On the other hand, the secondary interactions (i.e. dispersion, and hydrogen bonds) between the substituents could result in a LA/LB complex. Such complex gains a name of the Frustrated Lewis pair, FLP, which was initially proposed by Stephan and his co-workers[2]. Interestingly, the first example of FLPs, $Mes_2P(C_6F_4)B(C_6F_5)_2$, was shown to activate hydrogen molecules ($H_2$) reversibly[2]. The author found that injection of $H_2$ into the solution of $Mes_2P(C_6F_4)B(C_6F_5)_2$ at 25 °C generated a product of $[Mes_2PH]^+(C_6F_4)[BH(C_6F_5)_2]^-$, and heating the product to 150 °C led to regeneration of $H_2$ and the phosphino-borane reactant. After that, a large number of inter-/intra-molecular FLPs have been explored[3-5]. Many important applications have been reported, such as activation of $H_2$[6], capture of $CO_2$[7], $SO_2$[8] and $NO_x$[9], and hydrogenation reduction of several unsaturated chemical bonds, including $C=C$[10], $C=O$[11], $C=N$[12], $C\equiv C$[13], $C=N$[14]. Nowadays, the FLPs chemistry have become one of the emerge fields because of their special electronic structures and important applications.[15]

The interesting application of FLPs towards $H_2$ activation have drawn great attention from the theoretical and computational chemists as well[16-18]. Density functional theory (DFT) and molecular dynamics (MD) simulations studies conclude that the formation of FLPs and its reactivity towards $H_2$ is strongly determined by the distance between LAs and LBs. Among the configurations of LA/LB complexes, only those with the correct orientation and suitable LA-LB distance can be considered as reactive pairs which activate small molecules[19]. According to the previous static DFT calculations on the $H_2$ activation by a series of FLPs[20-22], the distance between active centers of reactive FLPs ranges from 3 Å to 5 Å[23]. MD studies further expanded these distances to be no more than 6 Å by taking flexibility of FLPs into account[24-25]. However, MD simulations of $(tBu)_3P/B(C_6F_5)_3$ in toluene found that the amount of effective LA/LB pairs are rather low[25]. The author found that probability of finding pairs with LA-LB distance shorter than 6 Å is only ca. 2 % of the total simulated

LA/LB pairs, which decreases to be 0.5 % if the distance is limited to 5 Å.

In a recent study, Brown et al. [26] found that the population of effective FLPs has been significantly increased when replacing the toluene solvent with ionic liquids (ILs). In that work, the authors investigated the structures and interactions of $(t\text{Bu})_3\text{P}/\text{B}(\text{C}_6\text{F}_5)_3$ in the [$\text{C}_{10}$mim][NTf$_2$] IL by $^{19}$F and $^{31}$P NMR spectroscopy, and they found that the amount of borane-phosphine pairs which have P-B interactions has been increased to be more than 20%. Herein, we performed MD simulations to reveal the mechanism of interactions between ILs and FLPs, and its effects on the structures of FLPs, which is still unclear to the best of our knowledge. The results show that ILs are more intended to form a solvated shell surrounding the LA/LB pairs, which increases the amount of effective FLPs. Moreover, we find that such effects are somehow related to the alkyl chain length of the cations.

**Computational details**

All MD simulations were performed for the FLP ($\text{B}(\text{C}_6\text{F}_5)_3/(t\text{Bu})_3\text{P}$), toluene [$\text{C}_6$mim][NTf$_2$], [$\text{C}_{10}$mim][NTf$_2$] and [$\text{C}_{14}$mim][NTf$_2$] mixtures using the GROMACS 5.1.4 package[27]. The simulated systems contain 1000 pairs of $\text{B}(\text{C}_6\text{F}_5)_3/(t\text{Bu})_3\text{P}$ molecules and 7000 pairs of ILs in cubic cells, having size length of 140, 170, 240 and 400 Å for toluene, [$\text{C}_6$mim][NTf$_2$], [$\text{C}_{10}$mim][NTf$_2$] and [$\text{C}_{14}$mim][NTf$_2$], respectively. The packaging software uses packmol 18.169 package[28]. For the $(t\text{Bu})_3\text{P}$ and $\text{B}(\text{C}_6\text{F}_5)_3$ molecules, previous force field parameters have been utilized without any modification,[26] and atomic interactions were represented by OPLS force fileds[29-30]. For toluene, the OPLS-AA force field [29-30] was used. For the ILs ([$\text{C}_6$mim][NTf$_2$], [$\text{C}_{10}$mim][NTf$_2$] and [$\text{C}_{14}$mim][NTf$_2$]), both the force field parameters of ions and the point charges of atoms are taken from the previous fitted data.[31-32]

For all systems, energy minimization was first performed. Then an isothermal-isochoric (NVT) equilibration was carried out at a temperature of 300K, and an isothermal-isobaric (NPT) equilibration was performed at a pressure of 1 bar and at a temperature of 300 K. In MD simulations, the integration step was set to be 2 fs. The production runs were performed for 50 ns in NPT ensemble. The equation of motion was integrated by leap-frog algorithms[33]. The temperature was kept constant using the Nose-Hoover thermostat[34] with the relaxation times of 0.2 ps, and the pressure was maintained by using the Parrinello-Rahman barostat[35] algorithm with the coupling time of 2 ps. Bond lengths involving hydrogen atoms were kept constant by the LINCS algorithms[36]. All intermolecular interactions between atoms in the simulated systems were calculated within the cut-off distance of 1.2 nm, and the long-range electrostatics interaction was calculated by Particle Mesh Ewald method[37]. The radial distribution function (RDF) was obtained from the gmx-rdf module of GROMACS software.[27] The structures are presented by the visual molecular dynamics (VMD) package.[38]

**Results and Discussion**

At first, the force fields parameters were validated by comparing the simulated density of ILs with the experimental data. **Table 1** summarizes the density of toluene, [$C_6$mim][$NTf_2$], [$C_{10}$mim][$NTf_2$] and [$C_{14}$mim][$NTf_2$] from MD simulations and experiments. As we can see that the simulated density of toluene perfectly fits into the experimental value with almost no derivation (0.02 %). Moreover, the simulated densities of three ILs are also in good agreement with the experimental data, with relative errors being 1.19%, 1.41% and 2.50% for [$C_6$mim] [$NTf_2$], [$C_{10}$mim][$NTf_2$] and [$C_{14}$mim][$NTf_2$], respectively. Those small errors indicate that our force fields parameters and the settings in MD simulations are accurate enough to describe the interactions of the studied systems.

**Table 1**. The simulated and experimental densities of toluene, [$C_6$mim][$NTf_2$], [$C_{10}$mim][$NTf_2$] and [$C_{14}$mim][$NTf_2$].

| Systems | Simulated data ($kg/m^3$, 300 K) | Experimental data ($kg/m^3$, 298.15 K) | Errors in % |
|---|---|---|---|
| toluene | 867.0 | 862.2[39] | 0.6 |
| [$C_6$mim][$NTf_2$] | 1386.3 | 1372.1[40] | 1.2 |
| [C$_{10}$mim][NTf$_2$] | 1294.0 | 1278.0[41] | 1.4 |
| [C$_{14}$mim][NTf$_2$] | 1232.1 | 1201.0[42] | 2.5 |

The MD simulations were first performed for the systems consisting of 1000 (*t*Bu)$_3$P/B(C$_6$F$_5$)$_3$ and 7000 toluene, 7000 [C$_{10}$mim][NTf$_2$], respectively. The two systems were firs equilibrated for 50 ns, and the continued simulations were run for another 50 ns. Based on these trajectories, we plot the RDF curves by choosing B atom in B(C$_6$F$_5$)$_3$ as the reference, which are depicted in **Fig. 2**. In the case of toluene (**Fig. 2a**), the firs peak of (*t*Bu)$_3$P (black line) appears around 5.5 Å, the second peaks appears around 6.5 Å, and the third one appears about 8.0 Å. Moreover, we found that the first peak of the toluene (red line) appear around 6.0 Å, which is very close to first peak of the (*t*Bu)$_3$P. Note that the height of such peak for toluene is almost four times larger than that of (*t*Bu)$_3$P. These findings indicate that in this region, toluene competes with (*t*Bu)$_3$P to interact with B(C$_6$F$_5$)$_3$, and somehow have stronger interactions with B(C$_6$F$_5$)$_3$, resulting in a rather low population of effective (*t*Bu)$_3$P/B(C$_6$F$_5$)$_3$ pairs. In the case of [C$_{10}$mim][NTf$_2$], the RDF profile of (*t*Bu)$_3$P does not show significant difference to that in toluene. For example, three peaks have been found in the positions of ca. 5.5, 6.5 and 8.0 Å, respectively. Their relative heights remain almost the same compared to the situation in toluene, where large population of (*t*Bu)$_3$P/B(C$_6$F$_5$)$_3$ pairs lies in the region with P-B distance larger than 6.0 Å. Interestingly, heights of the peaks at around 6.0 Å for [C$_{10}$mim]$^+$ and [NTf$_2$]$^-$ decrease dramatically, and apparent peaks appear around 8.0 Å, which is larger than the case of toluene (6.0 Å). These findings indicate that toluene molecule tends to separate the (*t*Bu)$_3$P/B(C$_6$F$_5$)$_3$ complex, while [C$_{10}$mim][NTf$_2$] prefer to form a solvation shell which surrounds the (*t*Bu)$_3$P/B(C$_6$F$_5$)$_3$ pairs.

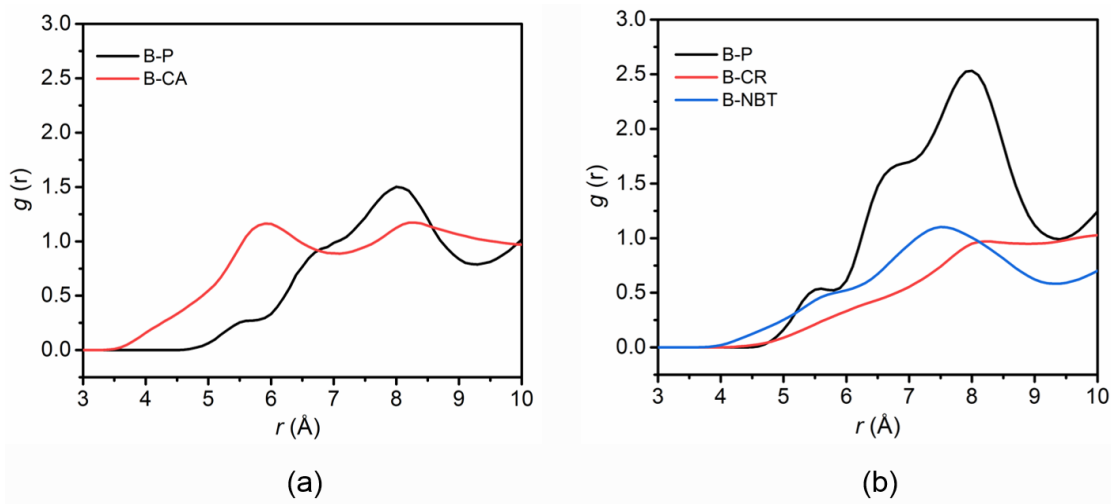

**Figure 2**. Computed RDF for $(tBu)_3P/B(C_6F_5)_3$ in toluene (a), and in $[C_{10}mim][NTf_2]$ (b). Notation: CA: $CH_3$ connected C atom in toluene, CR: $C_2$ atoms of $C_{10}mim$, and NBT: N atom of $NTf_2$.

Next, we analyzed the spatial distribution function (SDF) of the simulated systems, which gives the probability of atoms or molecules appearing in the 3D space around the central molecule. Visualization of SDF was performed using the gOpenMol 3. 00 software[43-44], and the results are given in **Fig. 3a** and **Fig. 3b** for solvents of toluene and $[C_{10}mim][NTf_2]$, respectively. In the case of toluene, SDF shows that toluene molecules forms an almost closed surface (blue) around the central $B(C_6F_5)_3$ molecules, and somehow separate them from the $(tBu_3)P$ molecules. Hence, most of the $(tBu_3)P$ molecules (red surface) distribute far from away from $B(C_6F_5)_3$ molecules. When having $[C_{10}mim][NTf_2]$ as the solvent, the SDF shows a rather different profile (**Fig. 2b**): $[C_{10}mim]$ (pink) and $[NTf_2]$ ions (yellow) also distribute around the central $B(C_6F_5)_3$ molecules. However, they leave relative large empty spaces which are then occupied by $(tBu_3)P$ molecules (red), resulting in higher possibilities of $B(C_6F_5)_3/(tBu_3)P$ interactions, and higher contents of effective FLPs. Subsequently, we selected two represented structures with P-B distance less than 4.8 Å for the solvents of toluene (**Fig. 3c**) and $[C_{10}mim][NTf_2]$ (**Fig. 3d**), respectively. In both cases, the solvents molecules form a solvation-shell as the outer layer, and the $B(C_6F_5)_3/(tBu_3)P$ pairs shows typical features of FLPs, that is, the Lewis acid and base centers face to each with P-B distances being ca.4.8 Å. As pointed out by previous DFT and MD studies[45], these two configuration are the ones responding for $H_2$ activation. Moreover, we found that the [$C_{10}$mim][$NTf_2$] molecules pack more condense than the toluene around the $B(C_6F_5)_3$/($t$Bu$_3$)P. On the one hand, the density of toluene is much smaller than that of [$C_{10}$mim][$NTf_2$] (862.2 kg/m$^3$ versus 1278.0 kg/m$^3$). On the other hand, the interactions between [$C_{10}$mim]$^+$ and [$NTf_2$]$^-$ are estimated to be -88.3 kcal mol$^{-1}$, while interactions between toluene molecules are -5.9 kcal mol$^{-1}$ at the B3LYP-D3/6-311+G* level of theory. These might be the reason that more effective $B(C_6F_5)_3$/($t$Bu$_3$)P pairs have been found in the case of [$C_{10}$mim][$NTf_2$] than that in toluene.

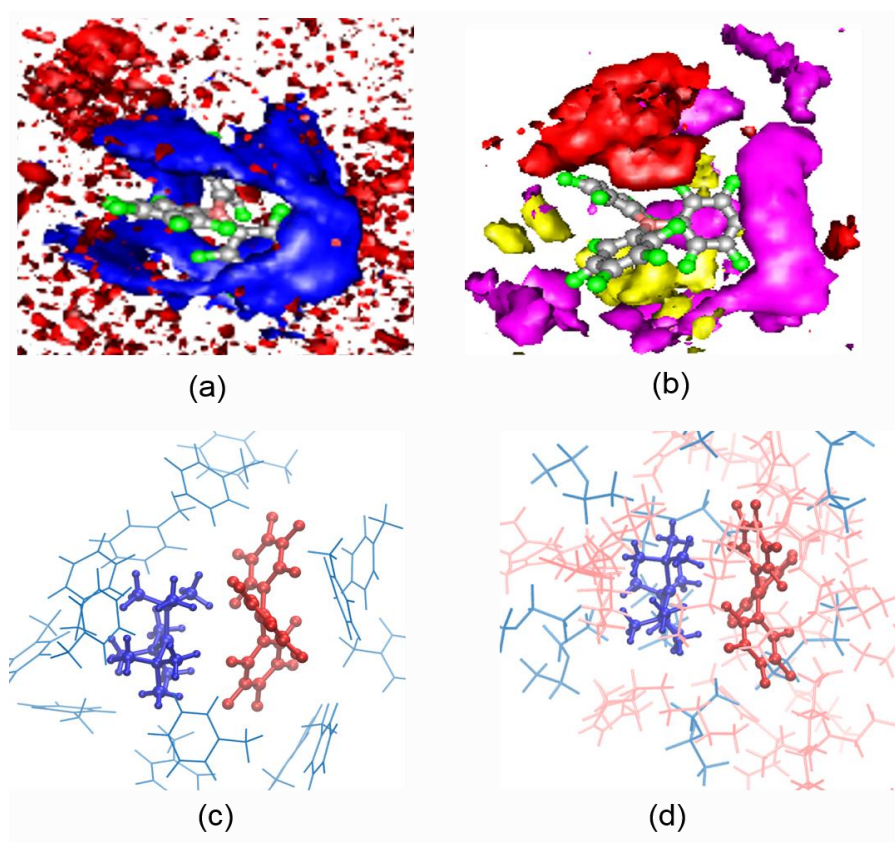

**Figure 3**: Computed RDF for ($t$Bu)$_3$P/$B(C_6F_5)_3$ in toluene (a), and in [$C_{10}$mim][$NTf_2$] (b) with $B(C_6F_5)_3$ as the central molecule. Color legend: toluene blue, ($t$Bu)$_3$P red, [$C_{10}$mim] pink, and [$NTf_2$] yellow. Selected structures with the P-B distance smaller than 5.0 Å in toluene (c), and in [$C_{10}$mim][$NTf_2$] (d). Color legend: $B(C_6F_5)_3$ dark red, ($t$Bu)$_3$P dark blue, [$C_{10}$mim] light red, and toluene/[$NTf_2$] light blue.

Lastly, we performed MD simulations to investigate effects of length of alkyl chain in cations on the structures of FLPs. 1000 $B(C_6F_5)_3/(tBu_3)P$ pairs were mixed with 7000 $[C_6mim][NTf_2]$ and 7000 $[C_{14}mim][NTf_2]$ in a cubic box with the length of 170 and 400 Å, respectively. Similar to that of $[C_{10}mim][NTf_2]$, both systems were equilibrated for 50 ns, and another 50 ns NPT runs were performed for the *post*-analysis. **Fig. 4** summarizes the computed RDF of $B(C_6F_5)_3/(tBu_3)P$ and $[C_6mim]$ $[NTf_2]$, $[C_{10}mim][NTf_2]$ and $[C_{14}mim][NTf_2]$. Generally, the shape of the RDFs are rather similar. For the interactions between $B(C_6F_5)_3$ and $(tBu_3)P$ ($C_n$-B-P), we found three peaks at the distance of 5.5, 6.5 and 8.0 Å, respectively. Both heights of the peaks at 6.5 Å and 8.0 Å are much larger than that of the peak at 5.5 Å. These findings indicate that most of the $B(C_6F_5)_3$ and $(tBu_3)P$ complexes have the P-B distance larger than 6.0 Å, and the amount of effective $B(C_6F_5)_3/(tBu_3)P$ pairs is relative small in all studied ILs. For the interactions between $B(C_6F_5)_3$ and ILs, we found two groups of peaks at around 6.0 Å and 8.0 Å, respectively, and the peaks corresponding to the interactions between $B(C_6F_5)_3$ and $[NTf_2]^-$ (Cn-B-NBT) appear slightly before the ones describing the interactions between $B(C_6F_5)_3$ and $[C_nmim]^+$ (Cn-B-CR). For example, two peaks of Cn-B-NBT curves have been found at 5.5 Å and 7.5 Å, while two peaks of Cn-B-CR curves have been found at 6.0 Å and 8.0 Å. As such, we conclude that $B(C_6F_5)_3$ interacts stronger with $[NTf_2]^-$ than $[C_nmim]^+$.

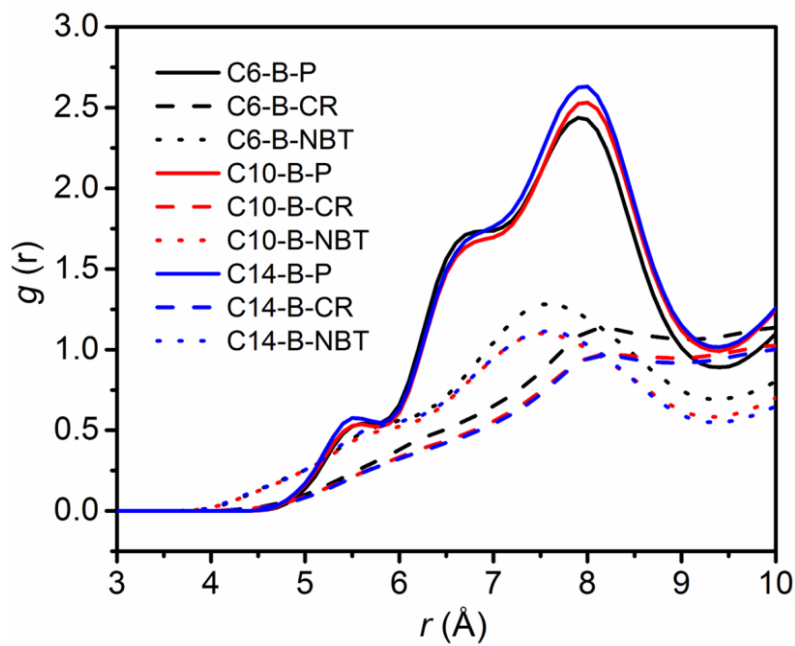

**Figure 4**: Computed RDF for $(tBu)_3P/B(C_6F_5)_3$ in $[C_nmim]$ ($n$ = 6, 10, 14). Notation: CR: $C_2$ atoms of $C_nmim$, and NBT: N atom of $NTf_2$.

Typically, the $g(r)$ function describes the distribution of the $B(C_6F_5)_3/(tBu_3)P$ pairs along the P-B distances ($d_{PB}$), and integration over $r$ gives the probability of finding LA/LB pairs in these states, $N(r)$. We then computed the $N(r)$ for three representative regions similar to the previous studies (**Table 3**)[25]: 1) $d_{PB} \leq 6.0$ Å, in which the LAs/LBs associates into reactive FLPs for $H_2$ activation; 2) $6.0$ Å $< d \leq 7.0$ Å, in which LAs/LBs are still in contact, and partial spaces between LA and LB are accessible by solvent molecules; 3) $d_{PB} > 7.0$ Å, in which LA and LB are completely separated by the solvent molecules. First of all, we can clearly see that the reactive FLPs ($d \leq 6.0$ Å) is rather low in all studied systems, and most of the pairs are separated by the solvent molecules, as already indicated by the SDF profiles (**Fig. 2** and **Fig.4**). In the case of toluene, the $N(r)$ of $d > 7.0$ Å is 90.36 %, while the $N(r)$ of $d \leq 6.0$ Å is only 2.32 %. In other words, only 2.32 % amount $(tBu)_3P$ and $B(C_6F_5)_3$ associate into the so-called FLPs, which is rather similar to that reported by the previous MD study[25]. Secondly, all investigated ILs show positive effects on the population of $B(C_6F_5)_3/(tBu_3)P$ in the region of $d \leq 6.0$ Å. For example, when replacing the toluene with $[C_6mim][NTf_2]$, $N(r)$ of $d > 7.0$ Å decreases from 90.36 % to 81.24 % while $N(r)$ of $d \leq 6.0$ Å increases from 2.32 % to 4.75 %. Note that for [$C_{10}$mim][NTf$_2$], if we sum the $N(r)$ of $d \leq 6.0$ Å (where typical associated FLPs are formed), and the 6.0 Å $< d \leq 7.0$ Å (where partial associated FLPs are formed), a value of 18.1 % has been obtained. Such finding is consistent with the pervious experimental NMR study, where populations of B(C$_6$F$_5$)$_3$/(tBu$_3$)P pairs having P-B interactions is about 20 %[26]. Lastly, we found that $N(r)$ of $d \leq 6.0$ Å gradually increases from 4.75 % to 5.15 % when alkyl chain length in [C$_n$mim] cations increases from 6 to 14. Previous studies pointed out that ILs form self-aggregation structures (i.e. vesicle)[46]. Hence, we assume that the ILs with longer alkyl chain length might form larger vesicle structures, and more B(C$_6$F$_5$)$_3$/(tBu$_3$)P pairs could be accommodated in the inner cave. As a result, larger $N(r)$ values have been obtained for the ILs with longer alkyl chain length (i.e. [C$_{14}$mim][NTf$_2$]).

**Table 3**. The computed the $N(r)$ of B(C$_6$F$_5$)$_3$/(tBu$_3$)P in toluene, [C$_6$mim][NTf$_2$], [C$_{10}$mim][NTf$_2$] and [C$_{14}$mim][NTf$_2$] along the P-B distance.

|  | $d \leq 6.0$ Å | 6.0 Å $< d \leq 7.0$ Å | $d > 7.0$ Å |
|---|---|---|---|
| toluene | 2.32 % | 7.20 % | 90.36 % |
| [C$_6$mim][NTf$_2$] | 4.75 % | 14.01 % | 81.24 % |
| [C$_{10}$mim][NTf$_2$] | 4.80 % | 13.29 % | 81.91 % |
| [C$_{14}$mim][NTf$_2$] | 5.15 % | 13.58 % | 81.27 % |

**Conclusions**

In this work, we performed a series of molecular dynamics (MD) simulations to investigate the solvent effects of toluene and [C$_n$mim][NTf$_2$] on the structures of B(C$_6$F$_5$)$_3$/(tBu$_3$)P, and the MD results were further detailed analyzed by the radial distribution functions (RDF) and spatial distribution functions (SDF). The results show that the amount of reactive B(C$_6$F$_5$)$_3$/(tBu$_3$)P pairs ($d_{PB} \leq 6.0$ Å) are rather low in both studied solvents, and the most of the LA/LB pairs distribute in the region with P-B distance larger than 7.0 Å. However, we found that the populations of effective

FLPs has been enhanced when replacing the toluene by the $[C_nmim][NTf_2]$ ionic liquids from ca. 2 % to 5 %, which can be further enhanced to ca. 20 %, if the partial reactive FLPs (6.0 Å < $d$ ≤ 7.0 Å) are taken into account. Based on the structural analysis, we assume that such findings are because of the strong interactions between cations and anion, and the possible formation of vesicle structures. Moreover, we found that the enhancement of ILs on the FLPs structures is somehow relative to the electronic structures of the ILs, i.e. longer length of alkyl chain in cations ($[C_nmim]$, $n$ = 6, 12 and 14), increase more amount of the effective FLPs. Overall, we present an theoretical understanding of the effects of ILs on the structure of FLPs, which could help to develop ILs-based FLPs systems for important applications.

**Acknowledgement**

This work was financially supported by the National Natural Science Foundation of China (21978294, 21808224).

Unilamellar Vesicle Formation of Ionic Liquids in Aqueous Solutions. *Chemical Communications* **2013**, *49*, 5222-5224.